\pdfoutput=1
\documentclass[traditabstract]{aa} 
\usepackage[usenames,dvipsnames]{xcolor}
\usepackage{graphicx}
\usepackage{natbib}
\usepackage{ulem}
\usepackage{graphicx}
\usepackage{longtable}
\usepackage{subfig}
\usepackage{float}
\usepackage{txfonts}
\begin{document}

    \title{CFBDS J111807-064016: A new L/T transition brown dwarf
      in a binary system \thanks{Based on observations obtained 
  with MegaPrime/MegaCam, a joint
  project of CFHT and CEA/DAPNIA, at the Canada-France-Hawaii
  Telescope (CFHT) which is operated by the National Research Council
  (NRC) of Canada, the Institut National des Sciences de l'Univers of
  the Centre National de la Recherche Scientifique (CNRS) of France,
  and the University of Hawaii. This work is based in part on data
  products produced at TERAPIX and the Canadian Astronomy Data Centre
  as part of the Canada-France-Hawaii Telescope Legacy Survey, a
  collaborative project of NRC and CNRS. 
  Based on observations made with the
ESO New Technology Telescope at the La Silla Observatory
under programme ID 082.C-0506(A) 
and the ESO Very Large Telescope at Paranal Observatory under programme ID 385.C-0242(A).
Based on observations (director discretionary time) made with the Canada-France-Hawaii Telescope at Mauna Kea
Observatory.
}}
\titlerunning{CFBDS~1118: a new  L/T transition brown dwarf in a binary}
\authorrunning{Reyl\'e et al.}

   \author{C. Reyl\'e
          \inst{1}
          \and
          P. Delorme\inst{2}
          \and
          E. Artigau\inst{3}
          \and
          X. Delfosse\inst{2}
           \and
	L. Albert\inst{3}
           \and
	T. Forveille\inst{2}
           \and
	A. S. Rajpurohit\inst{1}
           \and
	F. Allard\inst{4}
           \and
	D. Homeier\inst{4}
           \and
	A. C. Robin\inst{1}
                   }

\institute{Institut UTINAM, CNRS UMR 6213, Observatoire des Sciences de l'Univers THETA Franche-Comt\'{e}-Bourgogne, Universit\'e de Franche Comt\'{e}, 
Observatoire de Besan\c{c}on, BP 1615, 25010 Besan\c{c}on Cedex, France\\
 \email{celine@obs-besancon.fr}
\and
UJF-Grenoble 1 / CNRS-INSU, Institut de Plan\'etologie et
d'Astrophysique de Grenoble (IPAG) UMR 5274, Grenoble, F-38041, France
\and
D\'epartement de physique and Observatoire du Mont M\'egantic, Universit\'e de Montr\'eal, C.P. 6128, Succursale Centre-Ville, Montr\'eal, QC H3C 3J7, Canada
\and
Centre de Recherche Astrophysique de Lyon, CNRS UMR 5574, Universit\'{e} de Lyon, \'{E}cole Normale Sup\'{e}rieure de Lyon, 46 all\'{e}e d'Italie,
69364 Lyon Cedex 07, France
             }

\date{Received ...; accepted ...}

 \abstract
{Stellar-substellar binary systems are quite rare, and provide
  interesting benchmarks. They constrain the complex physics of
  substellar atmospheres, because several physical parameters of 
  the substellar secondary can be fixed from the much better
  characterized main sequence primary. We report the discovery of \object{CFBDS
  J111807-064016}, a T2 brown dwarf companion to  \object{2MASS
  J111806.99-064007.8}, a low-mass M4.5-M5 star. The brown-dwarf was 
  identified from the Canada France Brown Dwarf Survey.
   
At a distance of 50-120\,pc, the 7.7$^{\prime\prime}$  angular separation
corresponds to projected separations of 390-900\,AU. The primary
displays no H$_\alpha$ emission, placing a lower limit on the age of
the system of about 6 Gyr. The kinematics is also consistent with
membership in the old thin disc. 

We obtained near-infrared spectra, which together with recent
atmosphere models allow us determine the effective temperature and
gravity of both components. We derived a system metallicity 
of [Fe/H]=-0.1$\pm0.1$ using metallicity sensitive absorption 
features in our medium-resolution $K_s$ spectrum of the primary.
From these parameters and the age constraint, evolutionary models
estimate masses of 0.10 to $0.15 M_\odot$ for the M dwarf, and 0.06 to $0.07 M_\odot$ for the T dwarf.

This system is a particularly valuable benchmark because the  
brown dwarf is an early T: the cloud-clearing that occurs at the 
L/T transition is very sensitive to gravity, metallicity, and 
detailed dust properties, and produces a large scatter in the 
colours. This T2 dwarf, with its metallicity measured from 
the primary and its mass and gravity much better constrained than 
those of younger early-Ts, will anchor our understanding of the 
colours of L/T transition brown dwarfs. It is also one of the most 
massive T dwarfs, just below the hydrogen-burning limit, and all
this makes it a prime probe a brown dwarf atmosphere and evolution models.
 }
 {} 
   \keywords{CFBDS J111807-064016 -- 2MASS
  J111806.99-064007.8 -- Stars: Low mass, brown dwarfs -- binaries: general}

   \maketitle
%

\section{Introduction}

Observational degeneracies between the influences of age,  metallicity 
and effective temperature, hinder our understanding of the physics of 
atmospheres in the brown dwarf and exoplanetary temperature range. 
This leads to ambiguities in mapping observational measurements to 
physical parameters (effective temperature, mass, gravity, 
metallicity) using theoretical models. Those are also
much less reliable than for stellar objects, in part because 
the physics of their clouds challenges our understanding.

Brown dwarfs-main sequence star binaries uniquely break most of this
degeneracy, because several parameters can be obtained from the
better understood primary. Such benchmarks systems are rare, 
however. When the two components are sufficiently separated to 
be studied separately, simple spectroscopic observations of the 
bright primary measure the metallicity of the system and constrain 
its age.  This greatly helps interpreting the spectrum of the 
brown dwarf, by fixing its age and composition. The photometric 
distance of the primary -- and thus the luminosity of both 
components -- is also less uncertain, and follow-up parallax 
measurements are more easily carried out. \cite{Saumon.2006}, 
studying the atmosphere of \object{Gl 570 D}, found the first evidence for
departures from local chemical equilibrium in the high atmosphere
of T dwarfs, due to the kinetics of nitrogen and carbon chemistry 
in the presence of vertical mixing.
 \cite{Leggett.2008} showed that then current models can only fit 
\object{HN~Peg~B} at the minimum admissible age for the HN~Peg system 
and simultaneously require significant vertical atmospheric 
mixing. \cite{Burningham.2009} used the \object{Wolf~940} 
system to conclude that model-derived temperatures of very late
T dwarfs are 10\% too warm. Detailed studies 
of T dwarfs companion to main sequence stars therefore constrain 
our understanding of T dwarfs in general. Moreover, how often
such systems occur and the distributions functions of their
orbital parameters constrains theories for brown dwarf formation  
\citep{Reipurth.2001,Bate.2003,Stamatellos.2009}.

To date twenty resolved T~dwarf-main sequence star 
systems are confirmed (Table~1). 
These rare systems constrain cool atmosphere physics regardless of 
the brown dwarf spectral type, but they are most valuable in 
temperature ranges that models poorly describe. 
Our understanding of T dwarf atmospheres is most challanged at 
the L-T transition, where cloud-clearing processes
dramatically change the shape of emerging spectrum. While cool 
atmosphere models perform relatively well in fully dusty atmospheres 
(late-M to late-L), in fully dust-free atmospheres (early-T), and 
in the mid-T dwarfs where low temperature condensates appear 
\citep{Morley.2012}, 
they do not currently describe the more complex physics at the 
L/T transition very well \citep[e.g.][]{Allard.2001,Helling.2008}.

The transition is observationally characterized by the
  near-infrared colours swinging from very red, in the late-L and 
  early-T dwarfs, to very blue, in the mid-T dwarfs. This change occurs 
  over a narrow effective temperature range, $\sim$ 1100 to 1400 K 
  \citep{Kirkpatrick.2000,Golimowski.2004,Vrba.2004,Dupuy.2012}. 
The detection of variablity in a few of these objects
\citep[e.g.][]{Artigau.2009,Radigan.2012} moreover suggests 
  that cloud coverage is spatially inhomogeneous and further complicates 
  the modeling and interpretation.

This transition is currently poorly constrained
observationally. Of the 20 resolved (in
seeing-limited observations) binaries of T~dwarf and main sequence star, only 5 (\object{Gl 337 D}, \object{HD 46588 B}, \object{HN Peg B},
\object{$\epsilon$ Ind Ba}, and our discovery) have an early-T secondary, 
the others harboring mid-late T companions.\\

\begin{table*}[ht]
\label{tabbin}
\caption{Known binary systems with a main-sequence primary and a T
  dwarf secondary. The listed physical separations are the projected 
  separations. The last column is the reference of the
    discovery paper.}

\begin{tabular}{lllllll}

\hline
Name&	\multicolumn{2}{c}{Sp. Type}&	Dist. &\multicolumn{2}{c}{Sep.}&	Ref.\\
&\multicolumn{2}{c}{\hrulefill}& &\multicolumn{2}{c}{\hrulefill}&	\\
&	Primary	&Secondary&	(pc) &(AU)&($^{\prime\prime}$)&	\\
\hline
Gl~337 CD&	K1	&	L8/T0	&20.5	&881 &43&		\cite{Wilson.2001,Burgasser.2005}\\
HD~46588 B&	F7	&	L9		&17.9 &1420 &79.2&\cite{Loutrel.2011}\\
$\epsilon$~Ind Ba&K5&	T1		&3.6	 &1459	&405&	\cite{Scholz.2003}\\
CFBDS~1118&	M4.5 &	T2		&92&	709 &7.7	&This paper\\
HN~Peg~B&	G0V&	T2.5		&18.4	&795 &43.2&	\cite{Luhman.2007}\\
HIP~38939 B & 	K4 	& 	T4.5 		&18.5 &1630 &88 &\cite{Deacon.2012b}\\
2MASS J03202839-0446358$^a$&	M8.5	&T5&25	&$<$8.3 &$<$0.33	       &\cite{Burgasser.2008}\\
SDSS J000649.16-085246.3 B$^a$&	M8.5		&T5	&30	&--- &---	       &\cite{Burgasser.2012}\\
HD~118865 B&	F5&		T5		&62.4	&9200	&148	&\cite{Burningham.2013}\\
LHS 2803 B &M4.5	&T5.5	&21	&1400	&67.6 &\cite{Deacon.2012}\\
$\epsilon$~Ind Bb&K5&	T6		&3.6	 &1459	&405&	\cite{Scholz.2003}\\
SCR~1845 B&	M8.5&	T6		&3.8	&4.5&1.2&		\cite{Biller.2006}\\
HIP~73786 B&	K5	&	T6.5		&18.6	&1260&68	&\cite{Scholz.2010}\\
G~204-39	&	M3	&	T6.5		&13.6	&2685&198	&\cite{Faherty.2010}\\
Gl~229 B&		M1	&	T7		&5.8	&45&7.8&		\cite{Nakajima.1995}\\
HIP~63510 C&	M0.5&	T7		&11.7	&1200&103	&\cite{Scholz.2010}\\
Gl~570 D&		K4	&	T7.5		&5.9	&1525 &258.3&		\cite{Burgasser.2000}\\
HD~3651 B&	K0	&	T7.5 		&11&480	&43.5&		\cite{Mugrauer.2006}\\
Ross~458 C&	M2	&	T8		&11.4	&1163&102&		\cite{Goldman.2010}\\
LHS~6176 B&	M4	&	T8		&18.7	&970&52&	\cite{	Burningham.2013}\\
Wolf~940 B&	M4	&	T8.5		&12.5	&400&32	&\cite{Burningham.2009}\\
GJ~758 B&		G9	&	T9		&15.5	&29&1.9&\cite{Thalmann.2009}\\
\hline
\end{tabular}

$a$ unresolved system
\end{table*}

We report here the discovery of \object{CFBDS J111807-064016} (hereafter
CFBDS~1118), an early T dwarf bound to a mid-M dwarf, 
\object{2MASS~J111806.99-064007.8} (hereafter 2MASS~1118). We
found the brown dwarf in the Canada France Brown Dwarf Survey 
(CFBDS), a wide field survey for cool brown dwarfs which we 
conducted with the MegaCam camera \citep{Boulade.2003proc} on 
the Canada France Hawaii Telescope (CFHT), and subsequently
identified the primary from its common proper motion.
In Section~\ref{obs} we describe our identification of the system
and its near-infrared photometric and 
spectroscopic follow-up. Section~\ref{phys} examines the physical 
properties of the system. Section~\ref{disc} discusses the 
physical properties of the T dwarf and contrasts them with 
those of \object{HN~Peg~B}, a T2.5 dwarf with very different age and mass. 
The last Section summarises our conclusions.


\section{Observations}
\label{obs}

We first identified \object{CFBDS~1118} as a brown dwarf candidate
from its red  $i^\prime-z^\prime$ colour in the CFBDS. The aims of the survey and its
detection techniques are fully described in \cite{Delorme.2008b}.
\object{CFBDS~1118} is undetected in $i^\prime$, with a $5\sigma$  
$i^\prime>24.8$ upper limit), while strongly detected at $z^\prime$ 
detection ($z^\prime=22.56{\pm}0.05$). The $i^\prime-z^\prime>2.3$
($5\sigma$) lower limit made \object{CFBDS~1118} a strong brown 
dwarf candidate.

\begin{figure}[h]
\begin{center}
\includegraphics[scale=0.6]{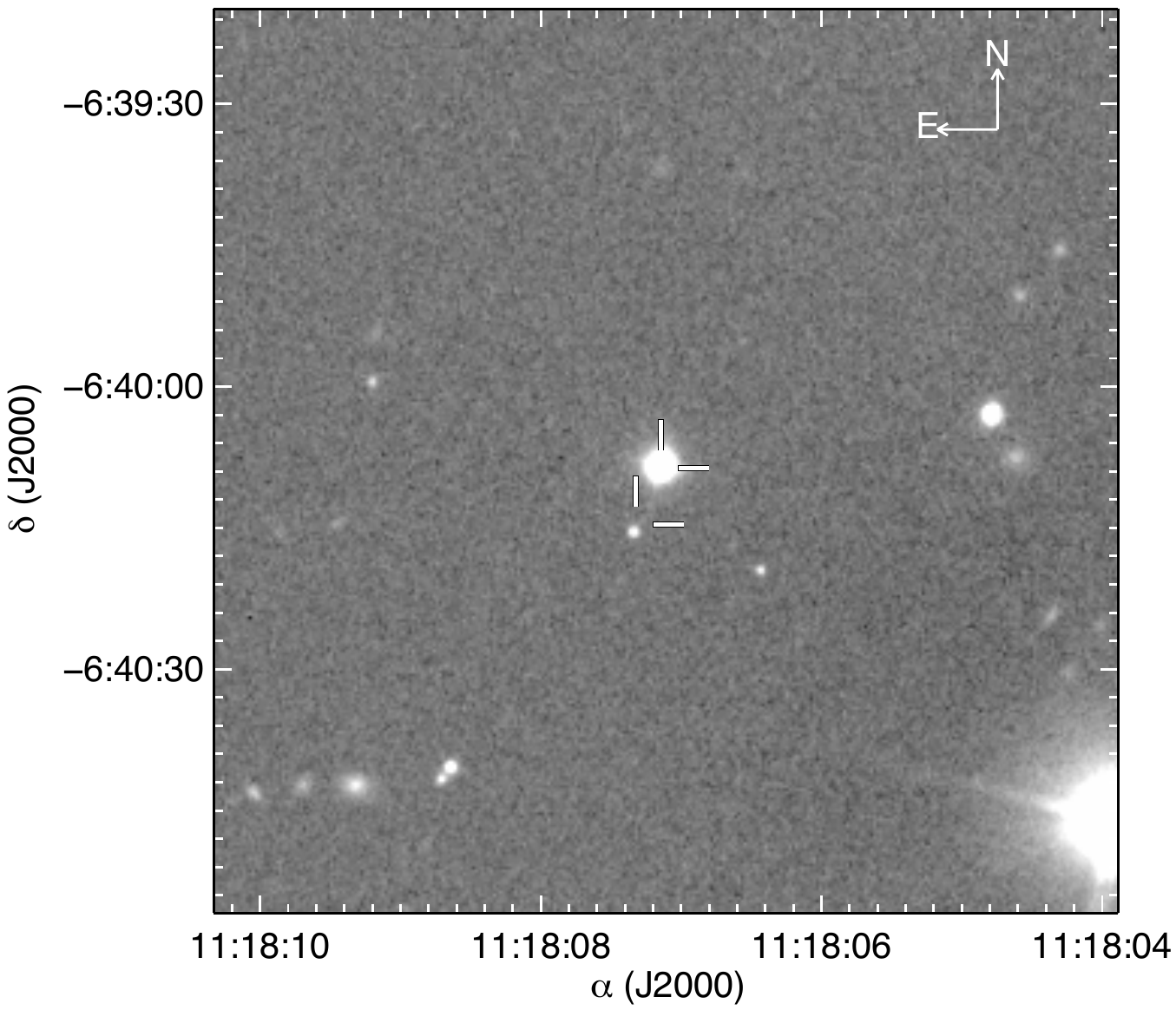}
\caption{$J-$band image obtained with WIRCam at CFHT. The tick
    marks identify both components, \object{2MASS~1118} (brighter) and 
    \object{CFBDS~1118} (fainter)}.
  \label{finder}
\end{center}
\end{figure}

Fig~\ref{finder} shows a $J$-band image of the brown dwarf, which 
lies 7.7$^{\prime\prime}$ off a much brighter star identified as
\object{2MASS~1118} 
\citep[in the 2MASS survey,][]{Skrutskie.2006}. 

The astrometry, proper motions, and photometry\footnote{Throughout
  this paper, we use Vega magnitudes for the near-infrared bands and
  AB magnitudes for the $i^\prime$ and $z^\prime$ optical bands.} of
the components are summarized in Table~2.

\begin{table}[h]
\label{tabphot}
\caption{Positions, proper motions, and photometry of both
  components. The photometry of the M dwarf is from 2MASS, and the 
  near-infrared  magnitudes of the T dwarf are on the 2MASS system.
  The $z^\prime$ magnitude is on the AB system used by CFHT
    for MegaCam\protect\footnotemark. 
    The proper motion of the T dwarf is computed between
    the first and third observation epoch. The proper motions are
  relative to the mean motion of the stars in the field.}
\center

\begin{tabular}{ccc}

\hline
&M dwarf &T dwarf\\
\hline
$\alpha$&11h18m06.99s&11h18m07.13s\\
$\delta$& $-06^\circ40^{\prime}07.84^{\prime\prime}$ &$-06^\circ40^{\prime}15.82^{\prime\prime}$\\
$i^\prime$ &&$>$ 24.86\\
$z^\prime$ &&$22.56\pm0.08$\\
$J$ &$13.84\pm0.03$&$19.01\pm0.02$\\
$H$ &$13.25\pm0.02$&$18.58\pm0.03$\\
$K_s$ &$12.95\pm0.03$&$18.30\pm0.03$\\
$\mu_\alpha$&$-201\pm11$ mas/yr&$-190\pm14$ mas yr$^{-1}$\\
$\mu_\delta$&$-49\pm20$ mas/yr&$-60\pm21$ mas yr$^{-1}$\\
\hline
\end{tabular}
\end{table}

\footnotetext{http://www.cfht.hawaii.edu/Instruments/Imaging/MegaPrime/ specsinformation.html}

\subsection{Near-infrared photometry}

We obtained near-infrared follow-up in the $J$, $H$, and $K_s-$bands
in the 2MASS system. Our $J-$band and $H-$band imaging 
respectively consist in twenty 30~seconds, and thirty 20~seconds 
long dithered exposures with the SOFI \citep{Moorwood.1998} 
near-infrared camera on the New Technology Telescope 
(NTT) at the European Southern Observatory (ESO), La Silla, 
on March 7, 2009 (Program 082.C-0506(A)). We used a modified
version of the jitter utility within the ESO Eclipse package
\citep{Devillard.1997} to correct for the flat field, subtract 
the background and coadd the exposures. We also obtained 
 $K_s-$band observations of \object{CFBDS~1118} with WIRCam 
\citep{Puget.2004} at CFHT on March 12, 2012. That sequence 
consists of 21, 20\,s dithered exposures. Table~3 summarizes
our imaging observations.

\begin{table}[ht]
\caption{Summary of our imaging observations of CFBDS~1118.}
\center

\begin{tabular}{cccc}

\hline
Filter & Camera & 	Exposures & 	date\\
\hline
$z^\prime $&  CFHT-MegaCam &  1x360s &  12 April 2007\\
$i^\prime $&  CFHT-MegaCam &  1X500s &  5 January 2008\\
$J $&  NTT-SOFI     &  20x30s &  5 March 2009\\
$H  $&  NTT-SOFI     &  31x20s &  7 March 2009\\
$K_s $& CFHT-WIRCam   &  17x21s &  12 March 2012\\
$J$  & CFHT-WIRCam   &   5x26s &  5 March 2012\\
\hline
\end{tabular}
\label{tablog}
\end{table}

We extracted photometry from the resulting images using
point spread function fitting within Source Extractor 
\citep{Bertin.1996} and obtain $J=19.01{\pm}0.02$, $H=18.58{\pm}0.03$, 
and $K_s=18.30{\pm}0.03$. The resulting $z^\prime-J=3.55{\pm}0.08$,  
$J-H=0.43{\pm}0.03$, and $H-K_s=0.38{\pm}0.04$ colours confirmed 
\object{CFBDS~1118} as a brown dwarf at the L/T transition 
\citep[see e.g. Figs. 1 in ][]{Reyle.2010,Delorme.2012}.
Besides confirming \object{CFBDS~1118} as a brown dwarf, the 
photometry will be used to flux-calibrate the spectrum (\S\ref{spectro}).

 \subsection{Proper motions}

The apparent magnitudes ($J=19.01$ and $J=13.84$) and colours of the M 
and T dwarfs places them in a common distance range, 70 to 120 pc from 
Earth (Sect.~\ref{dist}). Comparison of their proper motions 
further probes whether they form a physical binary. 

We first computed the proper motion of the T~dwarf from its 
positions in the $z^\prime-$band image (observed on April 12, 2007)
and in the $J-$band image (observed on March 07, 2009). That value
agreed within 1-$\sigma$ with the proper motion listed in the NOMAD 
Catalog \citep{Zacharias.2004} for the \hbox{M dwarf}, but 
was too noisy ($\sigma_{\mu_\alpha}=64$ mas yr$^{-1}$,
  $\sigma_{\mu_\delta}=97$ mas yr$^{-1}$) to strongly exclude a chance alignement.

We therefore obtained a third epoch observation in the $J-$band with
WIRCam at CFHT on March 5, 2012. We did not attempt to reference the astrometry to
an absolute frame, as experience shows that this results in larger
errorbars, and instead calculated a relative local astrometric
solution. This was achieved by cross matching the epochs and 
using Scamp \citep{Bertin.2006} to reference all astrometric
solutions to the first epoch image. Adding the third epoch 
reduces the error bars on the T dwarf proper motion measurements 
by a factor of 5, and allows a much stronger test for 
chance alignment. Table~2 lists those improved measurements.

 \subsection{Binarity characterization}

As Table~2 shows, the proper motions agree to within 1$\sigma$. 
Simulations of the galactic population with the Besan\c{c}on Galaxy 
model \citep{Robin.2003} in the direction of \object{CFBDS~1118} shows an 
only $\sim3\times10^{-5}$ probability that any main sequence star 
with a proper motion within 3$\sigma$ of the T dwarf lies by 
chance in the volume of space delimited by a cone section of 
8$\arcsec$ aperture and 120~pc depth in the direction of CFBDS~1118. 
Since the CFBDS survey found $\sim$70~T~dwarfs in CFBDS, the
probability that at least one of them would so closely align 
by chance with an unrelated main-sequence star is just
$\sim2\times10^{-3}$. Even one chance alignment within the 
full survey scale is thus very unlikely, and from thereon
we discuss \object{CFBDS~1118} and \object{2MASS~1118} as a physical binary system.

\subsection{Spectroscopic follow-up}
\label{spectro}

\begin{figure}[h]
\begin{center}
\includegraphics[scale=0.5,clip=]{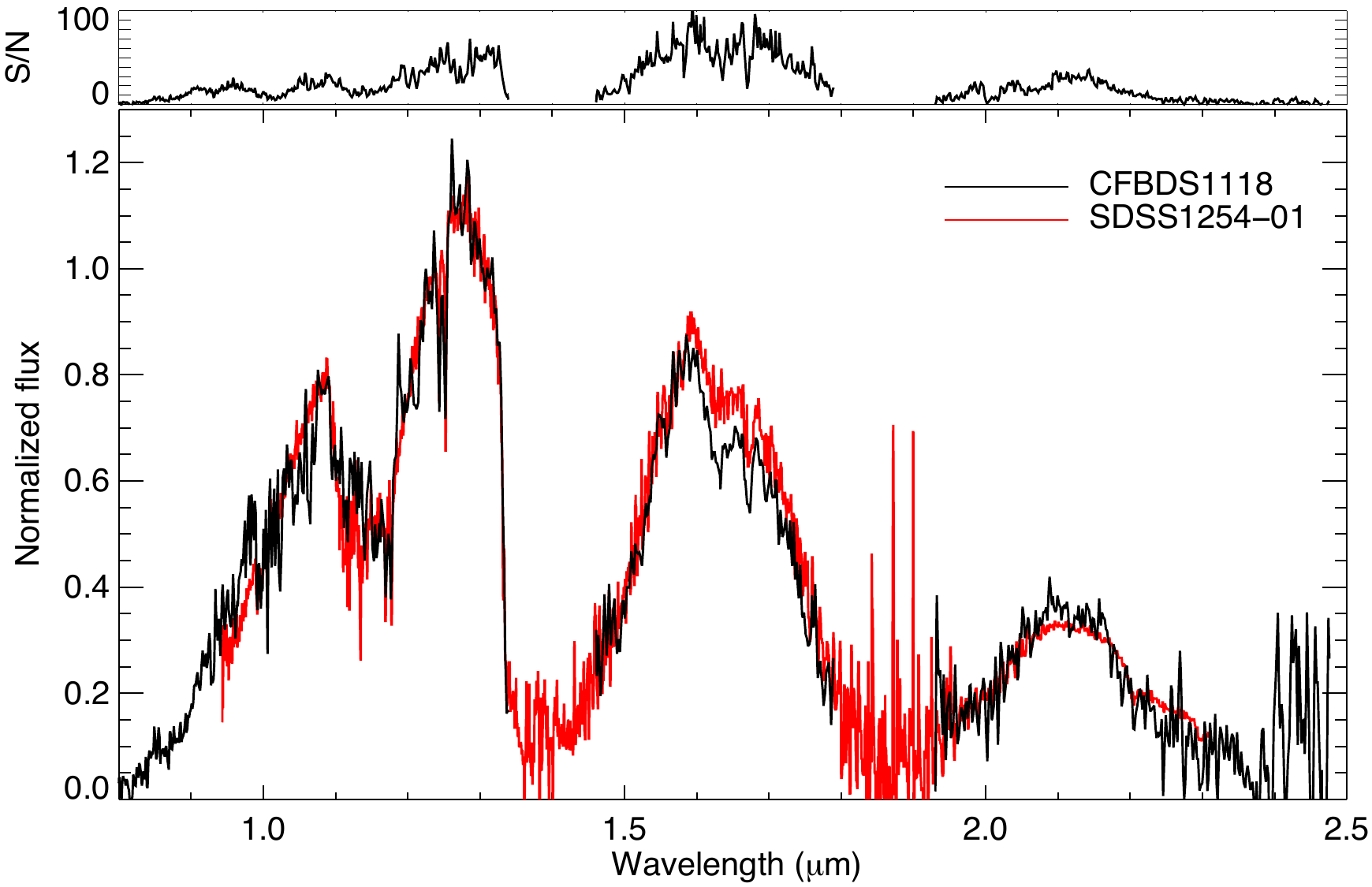}
\caption{
Black: spectrum of \object{CFBDS~1118} obtained with
  XSHOOTER at VLT. The regions with strong telluric H$_2$O absorptions cannot be used and are blanked out. Red: T2 dwarf, \object{SDSS~J125453.9-012247} \citep{McLean.2007}. A scaling factor is applied to normalize the median flux in the range 1.2$-$1.3 $\mu$m to unity. 
  The upper panel shows the signal to noise ratio.
\label{compT}
}
\end{center}
\end{figure}

We obtained near-infrared spectra of \object{CFBDS~1118} with the XSHOOTER 
spectrograph \citep{Vernet.2011} on the Very Large Telescope (VLT-UT2) at ESO through Program
385.C-0242(A). The observations were carried out as three 1-hour 
observing blocks on April 13, 2011, achieving a total exposure 
time on target of 132 minutes, split into 6 on-the-slit A-B nods 
of $2 \times660$\,s each. The slit width was 1.2$\arcsec$, and the 
seeing varied in the range 0.7-0.9$\arcsec$. We chose a wide slit,
at some cost in increased sky background, because the T~dwarf is 
not visible on the acquisition camera and was set on the slit 
through a blind offset from the primary. The spectra for 
individual observing blocks were reduced using the standard 
ESO XSHOOTER pipeline  \citep{Modigliani.2010},  which produces a 
2-dimensional,
curvature-corrected spectrum. We used our own IDL procedures to
extract the trace, using Gaussian boxes in the spatial dimension 
at each wavelength. A similar Gaussian extraction box 5~FWHM off
the trace was used to obtain the spectrum of the sky. The resulting
1-dimensional spectrum from each science target observing block 
was then divided by the spectra of reference stars observed 
just before or after in order to remove the telluric
  absorptions. Those were reduced and extracted using the same pipelines.
The resulting spectra from the three observing blocks were then 
median-combined into a final science spectrum. Since that spectrum 
(with $R=\lambda/\Delta\lambda\simeq$ 3900) has a low signal to noise 
ratio, we smoothed the spectra using a weighted average over 100
pixels in the wavelength dimension in the visible arm and over 20 pixels in the near-infrared arm. 
Our weighting is by the inverse 
variance of the extracted sky spectrum and therefore uses the 
full spectral resolution of XSHOOTER to downweight wavelengths 
affected by telluric emission lines. This improves the signal to 
noise ratio significantly over simple binning or observations with
lower resolution. The final resolution is $R\simeq$ 1100 and the resulting signal to noise ratio is shown in Figure~\ref{compT} (upper panel).\\

The same reduction and extraction procedures were
used for the near-infrared and visible arms of XSHOOTER. 
Given the wide wavelength range covered by XSHOOTER, from 0.6
  to 2.5 $\mu$m, seeing variation along the spectral direction can 
produce wavelength-dependent slit losses. We ensured the flux
homogeneity of this wide wavelength spectrum by calibrating it onto 
our NTT and WIRCam photometry in the $J$, $H$, and $K_s-$band 
(Table~2), and onto a near-infrared $Y-$band magnitude which we
estimated from the average colours of a T2 dwarf 
\citep{Burningham.2013}. Following the procedure described 
in \cite{Delorme.2010}, we applied small scaling factors to each 
broadband wavelength ranges to match synthetic photometry derived 
from the CFBDS~1118 spectrum to its observed photometry.
The final spectrum is shown in Figure~\ref{compT}.\\

\begin{figure}[ht]
\begin{center}
\includegraphics[scale=0.5,clip=]{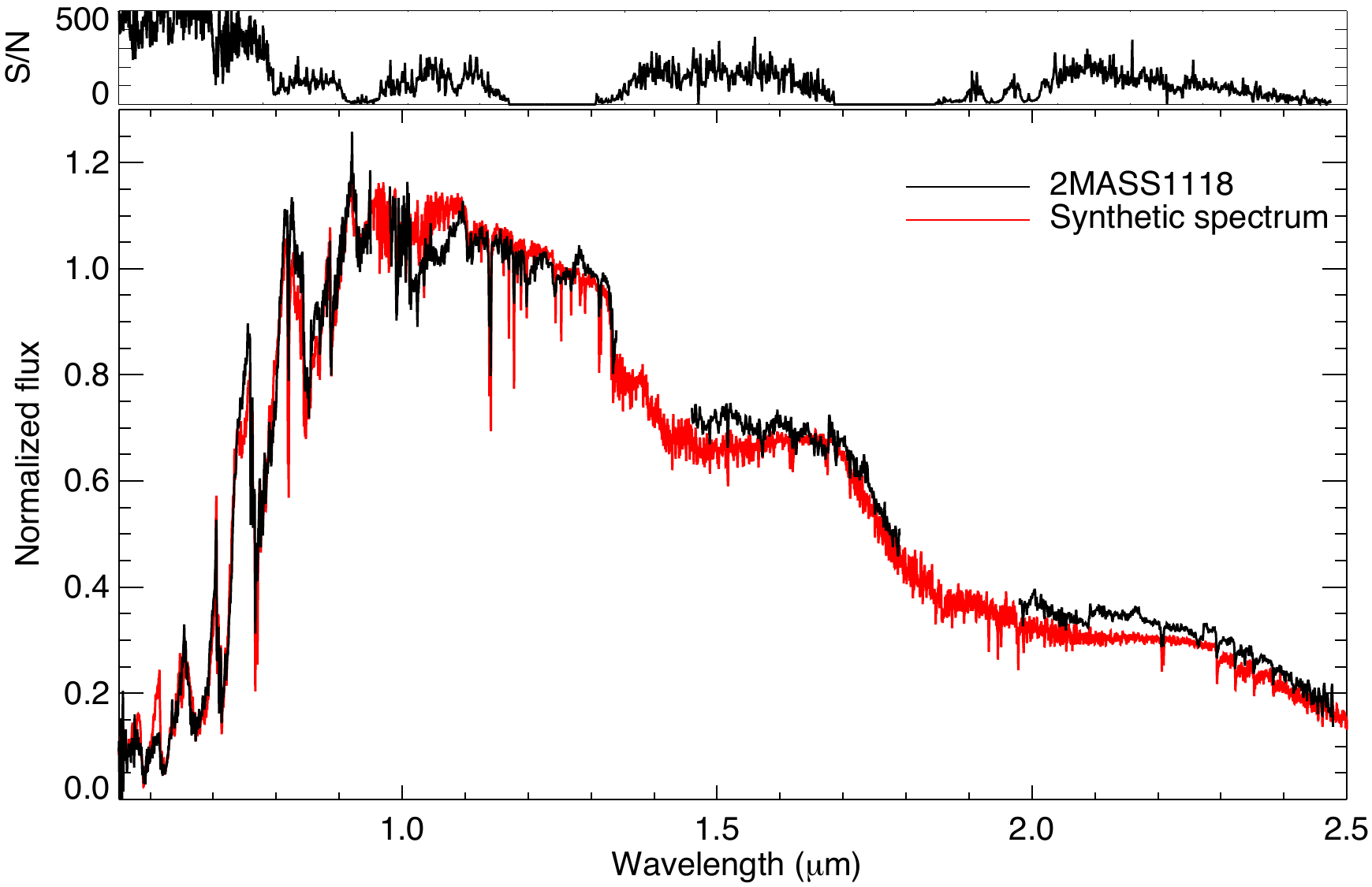}
\caption{
Spectrum of \object{2MASS~1118} obtained with XSHOOTER at VLT, in the visible
arm ($\lambda<1\mu$m) and the near-infrared arm ($\lambda>1\mu$m). The
regions with strong telluric H$_2$O absorptions cannot be
used and are blanked out. The red line shows the synthetic spectrum fitted to derive the astrophysical parameters (see \S\ref{phys}). A scaling factor is applied to normalize the median flux in the range 1.2$-$1.3 $\mu$m to unity.  The upper panel
  shows the signal to noise ratio.
\label{spectreMall}
}
\end{center}
\end{figure}

We also observed the primary star with XSHOOTER. The M dwarf is
detected with a high signal to noise ratio in all XSHOOTER arms 
(see upper panel in  Figure~\ref{spectreMall}) and its
spectrum was extracted following the procedures described for the T dwarf.
The final spectrum, smoothed using a weighted average over 10 pixels
in the wavelength dimension, is shown in Figure~\ref{spectreMall}.

\section{Physical properties}
\label{phys}

\subsection{Spectral type}
\label{dist}
We computed spectral indices and derived a spectral type for both
component. For the M dwarf, we used the classification scheme defined
by \cite{Reid.1995} based on TiO and CaH bands strengths. For the
T dwarf, we computed the spectroscopic indices defined by
\cite{Burgasser.2006,Delorme.2008a}, which quantify the strength of
the key molecular absorption bands, H$_2$O and CH$_4$. Table~4 lists
those spectral indices and the corresponding spectral types. This
classifies the primary as a M4.5 to M5 dwarf and \object{CFBDS~1118} as a T2 to
T3 dwarf. Figure~\ref{compT} shows the agreement between the spectra of \object{CFBDS~1118} and 
\object{SDSS~J125453.9-012247}, a T2 spectral type
  standard from the NIRSPEC Brown Dwarf Spectroscopic Survey \citep{McLean.2007}.

\begin{table}[h]
\label{tabphot}
\caption{Spectral indices and corresponding spectral type of the two components.}
\center

\begin{tabular}{ccc}

\hline
\multicolumn{3}{c}{M dwarf} \\
\hline
TiO 5& 0.265& M5\\
CaH 1& 0.707 & M4.5\\
CaH 2& 0.347 &M4.5\\
CaH 3& 0.638 &M5\\
\hline
\multicolumn{3}{c}{T dwarf} \\
\hline
H$_2$O-J& 0.472 &  T2\\
CH$_4$-J& 0.603&	T2\\
H$_2$O-H& 0.482&	T2\\
CH$_4$-H& 0.815&	T2\\
CH$_4$-K& 0.400&	T3\\
\hline
\end{tabular}
\end{table}

\subsection{Photometric distance and separation}

In order to compute the photometric distance of \object{2MASS~1118}, we used
several $M_J$ versus spectral type relations from the literature, 
described in \cite{Reyle.2006}. The $M_J$ absolute magnitude of a 
M4.5 to M5 dwarf ranges from 8.50 to 9.72, taking into account 
the uncertainties in the relation, and in particular the 
discontinuity in the relation at $M_J\simeq$8.5 (or 
spectral type M3.5/M4.5; \cite{Reid.2002paperI} discuss this
discontinuity in detail). This absolute magnitude range 
is consistent with $M_J=9.60$ for \cite{Baraffe.1998} evolution 
models with the stellar parameters estimated in forthcoming 
sections (\ref{age},\ref{metal}, and \ref{teff}). This translates
to a distance of 67 to 117 pc.

We estimate the distance to the T dwarf with the spectral type vs 
 $M_J$ relation derived  by \cite{Dupuy.2012} from brown dwarfs 
with a measured trigonometric parallax.
The absolute magnitude of a field T2 dwarf is $M_J=14.9\pm0.5$, translating
to a distance from 51 to 83 pc.

The components are separated by 7.7$^{\prime\prime}$. At the distance of 
the system, this translates to a projected separation of 393 to 901~AU.

\subsection{Kinematics and age}
\label{age}

The proper motion translates into a rather high tangential velocity 
($77\pm30$ km s$^{-1}$yr$^{-1}$), suggesting an old age. 
The M dwarf spectrum shows no obvious Doppler shift. At the resolution 
of XSHOOTER, this constrains the radial velocity to a  $-30$ to 
$30$ km s$^{-1}$ interval. From these kinematics, simulations with the 
Besan\c{c}on galactic population model \citep{Robin.2003} find 
81\% probability that the age is above 3~Gyr and 58\% probability that
it is above 5 Gyr (see Table~5). The older age derived from the
kinematics is independently corroborated by the absence of 
H$\alpha$ emission in the optical spectrum of the primary. 
According to \cite{West.2008}, such a low activity level for a mid-M 
dwarf indicates an age $\gtrsim$ 6 Gyr.

\begin{table}[h]
\label{BGM}
\center
\caption{Fraction of stars as a function of age simulated with the Besan\c{c}on Galaxy Model \citep{Robin.2003} in the direction of the system. The distance of simulated stars ranges from 60 to 120 pc,  their proper motion $\mu_\alpha$ from $-230$ to $-140$ mas/yr and $\mu_\delta$ from $-110$ to $10$ mas/yr, and their radial velocity from $-30$ to $30$ km s$^{-1}$. }
\begin{tabular}{ccc}
\hline
Population & Age &Fraction\\
\hline
Disc&$<$1Gyr	&3\% \\
&	1-3Gyr	&16\%\\
&	3-5Gyr	&23\%\\
&5-7Gyr  &21\%\\
&7-10Gyr  &24\%\\
Thick disc&	11Gyr	&13\%\\
Halo&	14Gyr &0\%\\
\hline
\end{tabular}
\end{table}

\subsection{Metallicity}
\label{metal}

\begin{figure}
\begin{center}
\includegraphics[scale=0.5,clip=]{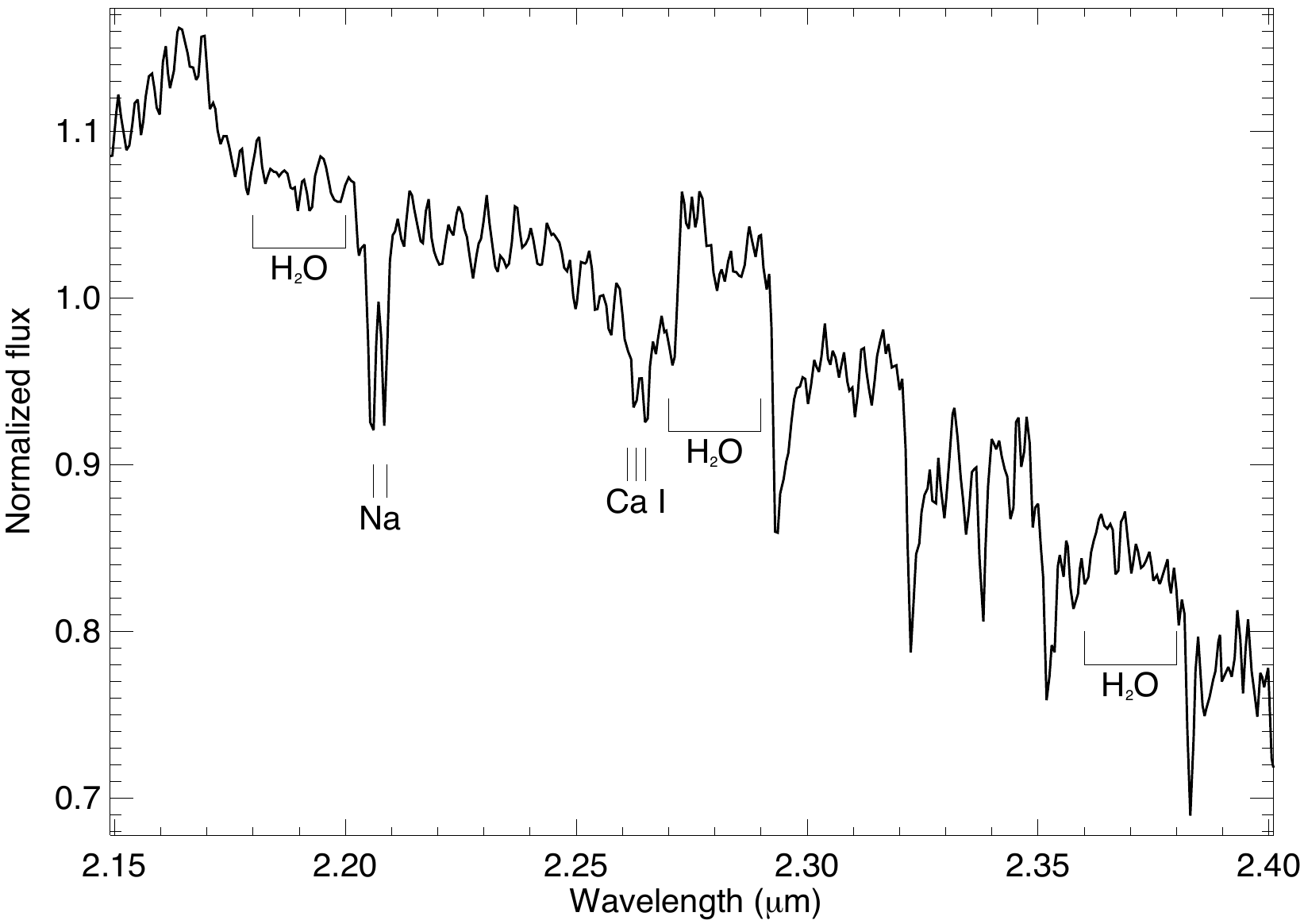}
\caption{
$K-$band 
spectrum of \object{2MASS~1118} obtained with XSHOOTER at VLT. The features used to compute the metallicity are highlighted.
\label{FeH}
}
\end{center}
\end{figure}

The metallicity of the system can be derived from the $K-$band
spectrum of the primary. As shown by \cite{Rojas-Ayala.2010}, the 
metallicity of M dwarfs can be inferred from the strength of 
their Na I (2.2~$\mu$m), Ca I (2.26$\mu$m) 
features, together with the 
H$_2$O-K index defined by Covey et al. (2010). 
These features are highlighted in Fig~\ref{FeH}.

From the \cite{Rojas-Ayala.2010,Terrien.2012} relations, the
metallicity is [Fe/H] = $-0.1\pm0.1$ and is consistent 
with the average metallicity in the solar neighborhood.

\subsection{Effective temperature, gravity, and mass}
\label{teff}

We have compared the most recent version of the BT-Settl stellar
atmosphere models \citep{Allard.2012,Allard.2013} with the observed
spectra of both components. Those models are computed with the 
PHOENIX code, and take into account : i) the solar abundances revised 
by \cite{Caffau.2011}, ii) the most recent BT2 version
of the water vapor line lists by Barber et al. (2008), 
iii) slightly revised atomic and molecular opacities, 
iv) a cloud model based on condensation and sedimentation timescales 
by Rossow (1978), supersaturation computed from pre-tabulated 
chemical equilibrium, and mixing from 2D radiation hydrodynamic
simulations by \cite{Freytag.2010}. The models are available
on-line\footnote{http://phoenix.ens-lyon.fr/simulator} and are 
fully described in \cite{Allard.2012,Rajpurohit.2012,Allard.2013}.

The M dwarf spectrum is best fitted by a synthetic spectrum with
$T_{\rm{eff}}=3000$ K $\mathrm{log}\,g = 5.0$ 
(Fig.~\ref{spectreMall}). The inferred temperature is consistent
with the relation between effective temperature scale and spectral 
sub-type for M dwarfs \citep[see e.g. Fig. 5 in][]{Rajpurohit.2013}. 
The \cite{Baraffe.1998} evolution models give 
$\mathrm{log}\,g = 5.2$ (and a mass of 0.10 to $0.15 M_\odot$)
for an age of 6 Gyr and $T_{\rm{eff}} = 3000 \pm  100$~K, which again
is reassuringly consistent with the value inferred from the spectrum.

\begin{figure*}[ht]
\begin{center}
\includegraphics[scale=0.9,clip=]{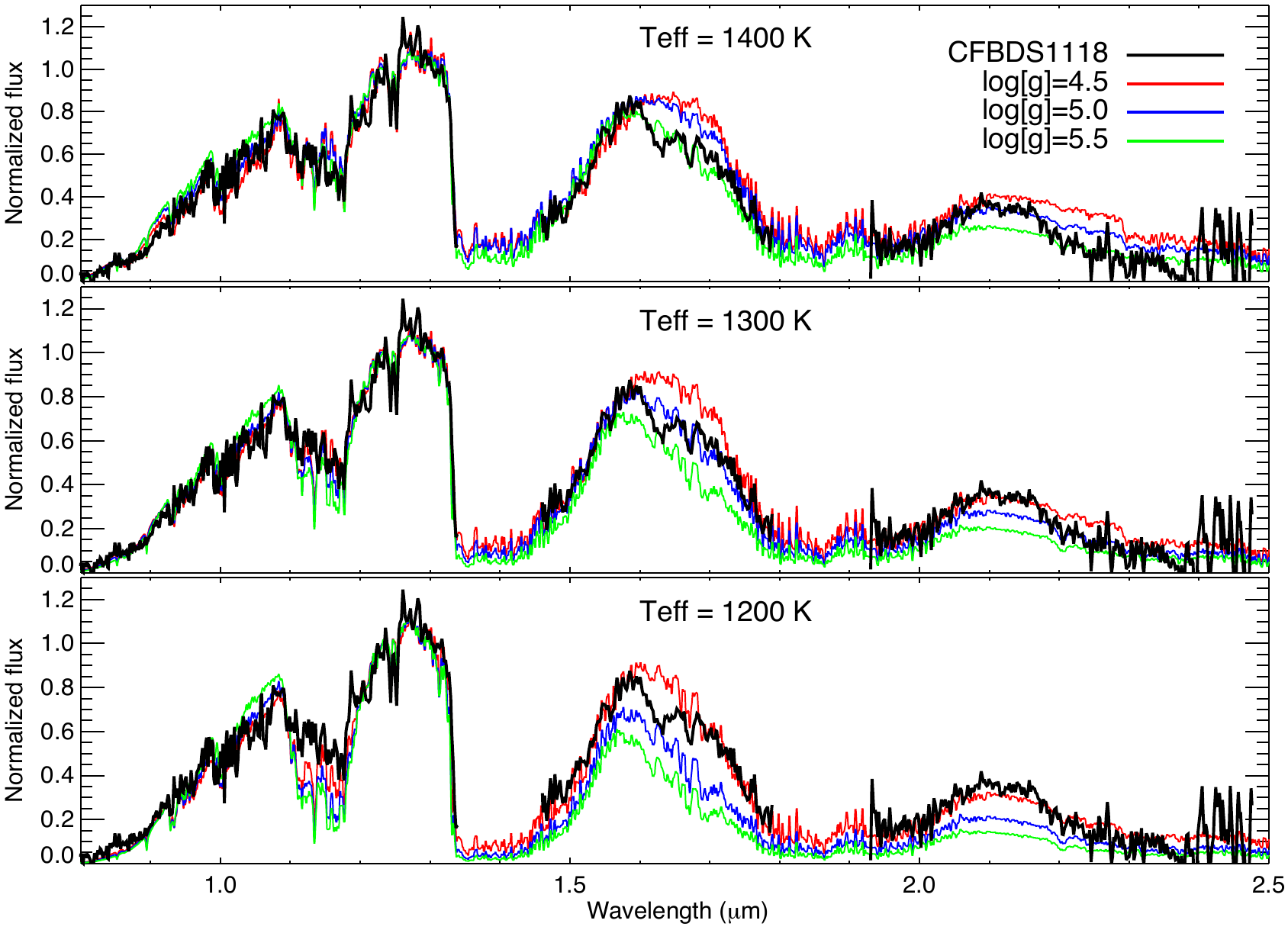}
\caption{Comparison of \object{CFBDS~1118} spectrum with BT-Settl models of varying effective temperature and
gravity. A scaling factor is applied to normalize the median flux in the range 1.2$-$1.3 $\mu$m to unity. The blue line in the middle panel shows the best fitting model 
($T_{\rm{eff}}=1300$ K and $\mathrm{log}\,g = 5.0$).
\label{modT}
}
\end{center}
\end{figure*}

Fig.~\ref{modT} compares the spectrum of the T dwarf with models of
varying effective temperature and gravity. We fixed the metallicity
of the models to solar, using our measurement from the near-infrared 
spectrum of the M dwarf to remove one degree of freedom. 
A scaling factor is applied to normalize the median flux in the range 1.2$-$1.3 $\mu$m to unity. 
The best fit, based on visual inspection of Fig.~\ref{modT},
is obtained for $T_{\rm{eff}}=1300$ K and $\mathrm{log}\,g = 5.0$. However one must be aware that the models still miss
some opacity in the $H-$band. Thus the 1200 K; 4.5 dex and 1400 K; 5.0
dex solutions cannot be excluded.

Table~6 compares these values to those obtained from the 
\citep{Baraffe.2003} evolution models, when adopting a
  $T_{\rm{eff}}=1300\pm100$ K and for a range of ages. For the
age range which we have adopted, the gravity value derived from 
the model atmosphere fit to the spectrum is inconsistent with that obtained from the evolution models at the fitted effective 
temperature.
This could potentially be due to 
(i)  an erroneous age estimate, or (ii) the known limitations of the cloud model, which is 
the largest source of uncertainty in brown dwarf atmosphere models. For example, the dust opacity might be poorly estimated due to an incomplete treatment of
grain growth mechanisms, which are being addressed in an upcoming revision of the models.
The evidence that 
the system is old, from the low magnetic activity level in the M dwarf
and the kinematics of the system, is strong.
This leaves (ii) as the most likely explanation.

\begin{table}[ht]
\center
\caption{Physical parameters derived from \cite{Baraffe.2003}
  evolution models. The listed temperatures are the closest 
  model grid points to our determination $T_{\rm{eff}}=1300\pm100$ K.}
\label{tab-baraffe}
\begin{tabular}{llllll}
\hline
Age &Mass & $T_{\rm{eff}}$ &$\mathrm{log}\,g$ &log L &Radius\\
(Gyr) &($M_\odot$) & (K) & & (L$_\odot$) & ($R_\odot$)\\
\hline
0.5& 0.03  &1264   &4.9 &-4.6    &0.1  \\
1 &0.04  &1271       &5.1 &-4.7  &0.09  \\
5 &0.06  &1120      &5.4  &-5.1    &0.08  \\
5 &0.07  &1524     &5.5  &-4.5    &0.08  \\
10 &0.07  &1289    &5.5   &-4.8   &0.08   \\
\hline
\end{tabular}
\end{table}

We therefore adopt the gravity range obtained from the evolution models at
old ages, $\mathrm{log}\,g = 5.4$ at 5 Gyr, and 5.5 at 10 Gyr.
This translates into a mass of 0.06 to 0.07 $M_\odot$ for \object{CFBDS~1118}, 
for ages of respectively 5 and 10~Gyr and a luminosity log L/L$_\odot = - 4.80 \pm 0.15$.

\section{Discussion}
\label{disc}

\begin{figure}[ht]
\begin{center}
\includegraphics[scale=0.45,clip=]{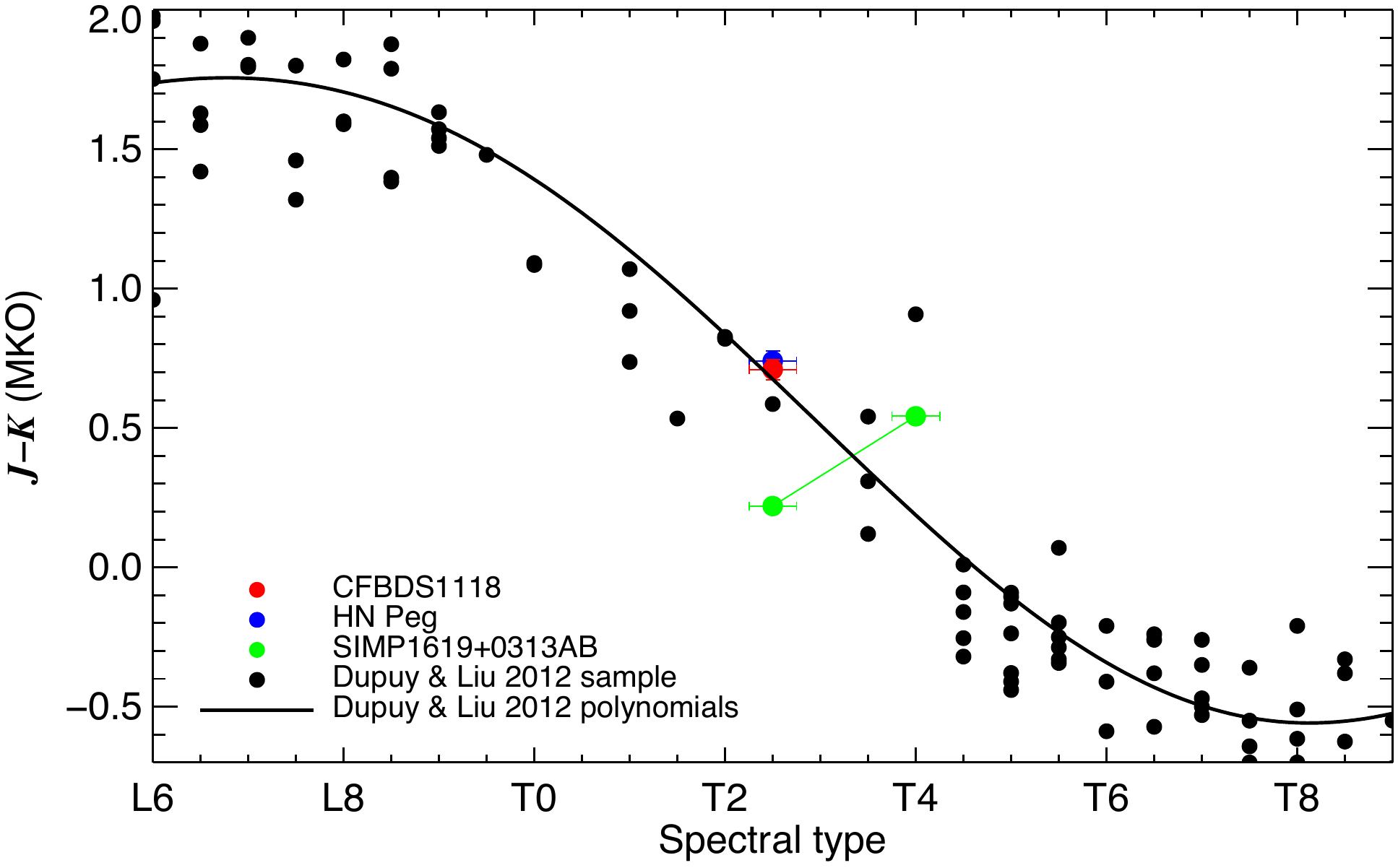}
\caption{$J-K$ (MKO) colour versus spectral type for L and T dwarfs. 
The red and blue markers highlight \object{CFBDS~1118} and \object{HN~Peg~B}, respectively.
The green symbols mark the two components of the \object{SIMP~1619+0313AB} binary.}
\label{colour-etienne}
\end{center}
\end{figure}

Fig.~\ref{colour-etienne} shows the colour versus spectral type
diagram for those L and T dwarfs with $J-K$ photometry measured to 
better than 0.25 mag and a near-infrared spectral type, as compiled by
\cite{Dupuy.2012}. The overall trend is a mild reddening of $J-K$ 
along the L-dwarf sequence, followed by its blueing along the T dwarf 
sequence. This behavior reflects the evolution of dust grains along
the L-dwarf sequence and their settling below the photosphere of 
T~dwarfs, as well as dispersion of the clouds and possibly 
other unknown mechanism. The diagram shows broad scatter at any 
given spectral type, most likely arising from differences in
metallicity, gravity, and detailed dust clouds properties. 
Redder colors can result from low gravity, high metallicity, more dust opacity, or a combination of these factors.

The diagram highlights \object{CFBDS~1118}, as well as \object{HN~Peg~B}. The latter
is a very young T2.5$\pm$0.5 dwarf, with an age of just
$0.3\pm0.2$\,Gyr estimated from its G0 main sequence primary. Its 
estimated mass is $0.021 \pm 0.009 M_\odot$, its effective temperature 
1130 $\pm70\,K$, and its gravity 4.8  
\citep{Luhman.2007,Leggett.2008}. The metallicity of the system 
is $-0.01 \pm 0.03$ \citep{Valenti.2005}.
These two T dwarfs therefore differ widely on age/mass/gravity, and slightly on effective temperature. 

Despite their contrasting ages and gravity, \object{CFBDS~1118} and \object{HN~Peg~B} 
have very similar $J-K$ colours, 0.71 and 0.74. Gravity therefore
contributes very little to the  $J-K$ scatter of early-T dwarfs, or its effect is compensated by other factors. 
As the L/T transition marks cloud-clearing, detailed dust properties
and cloud coverage may play an important role, as illustrated 
by the \object{SIMP~J1619+0313AB} T dwarf binary  \citep{Artigau.2011}. 
That system (also shown in Fig.~\ref{colour-etienne}) has 
unexpectedly reversed colours, with a bluer $J-K$ for the earlier-type 
(T2.5) primary than for the later-type (T4) secondary. The common 
age and metallicity of the two components leaves contrasting cloud 
properties, by elimination, as the preferred explanation for the
colour reversal \citep{Artigau.2011}.

\begin{figure}[ht]
\begin{center}
\includegraphics[scale=0.5]{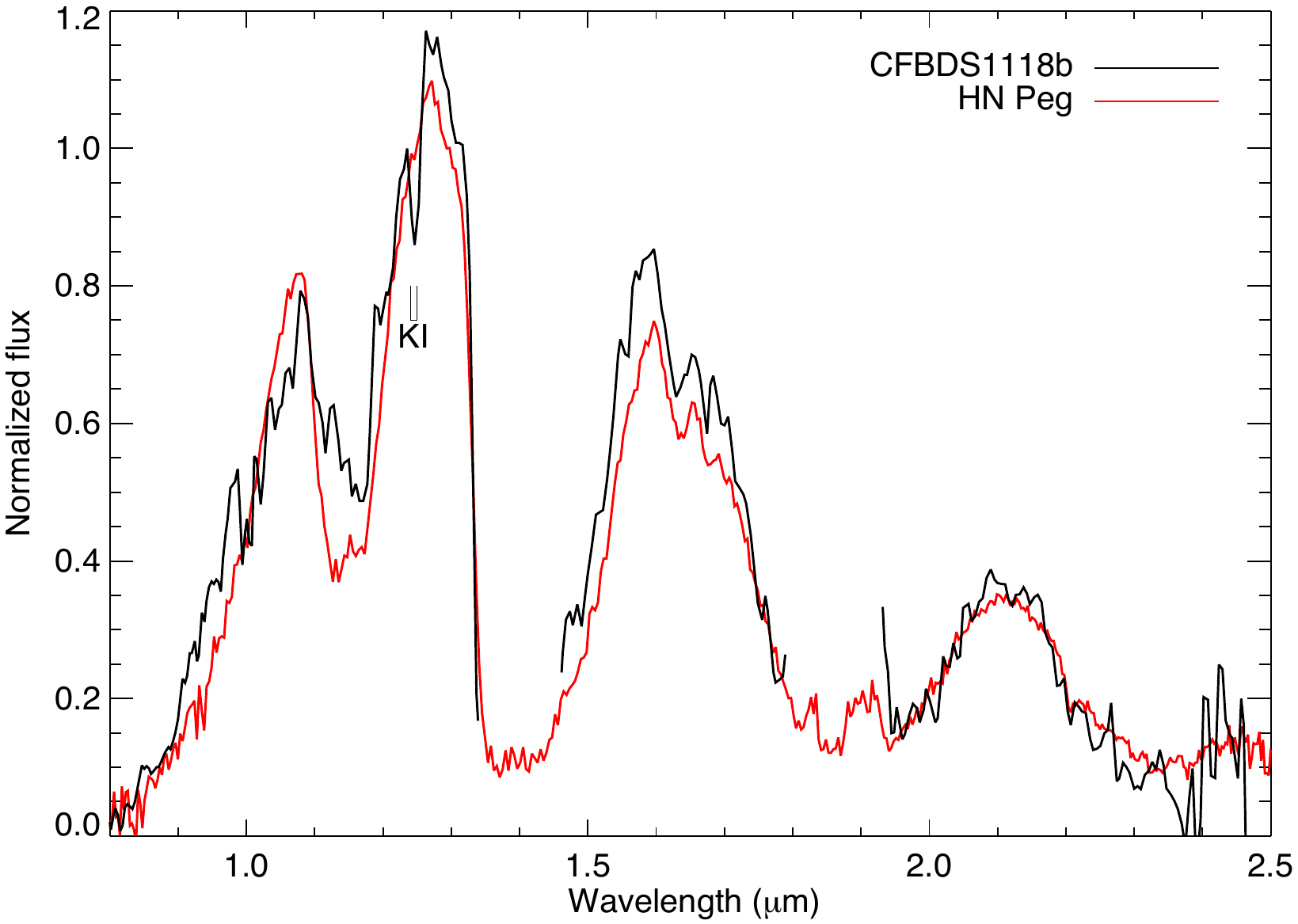}
\caption{
Spectrum of \object{CFBDS~1118} (black) obtained with XSHOOTER, binned to the
resolution of the  \citep{Luhman.2007} spectrum of \object{HN~Peg~B} (red) 
obtained with SpeX at the NASA Infrared Telescope Facility. A scaling factor is applied to normalize the median flux in the range 1.2$-$1.3 $\mu$m to unity. The 1.25$-$1.26 $\mu$m potassium doublet is highlighted.
\label{THNPeg}
}
\end{center}
\end{figure}

Fig.~\ref{THNPeg} compares the near-infrared spectra of \object{CFBDS~1118} and 
HN~Peg~B \citep[from][]{Luhman.2007}. The two spectra are 
extremely similar, with their potassium doublet feature as the main
difference. While the young, low-gravity, HN~Peg~B shows no
discernible K~I lines, that feature is readily apparent in the 
\object{CFBDS~1118} spectrum, as expected from its higher gravity (and older age). 
The increasing strength of K~I lines with increasing gravity 
has been extensively studied in L dwarfs \citep{Allers.2013}, but 
that trend (mildly) reverses for late-T dwarfs. Those have slightly 
stronger K I lines at lower gravity, because the CH$_4$ and H$_2$O 
bands which define their pseudo-continuum for the potassium doublet 
are slightly more sensitive to pressure broadening than is the atomic 
feature \citep{Allard.2012,Delorme.2012}. The two spectra have 
very similar (normalized) $K-$band flux, in spite of very contrasting 
ages and gravities. This constrasts with the late-T dwarfs, where high 
gravity enhances collision induced absorption by H$_2$  and,
for a fixed effective temperature therefore suppresses the 
$K-$band flux with respect  to the other bands 
\citep[e.g.][]{Knapp.2004,Delorme.2012}. The $J-K$ colour 
at the L/T transition is therefore controlled by some other 
parameter, which most likely is dust or cloud-related.

\begin{figure}[ht]
\begin{center}
\includegraphics[scale=0.47,clip=,bb=0 0 620 380]{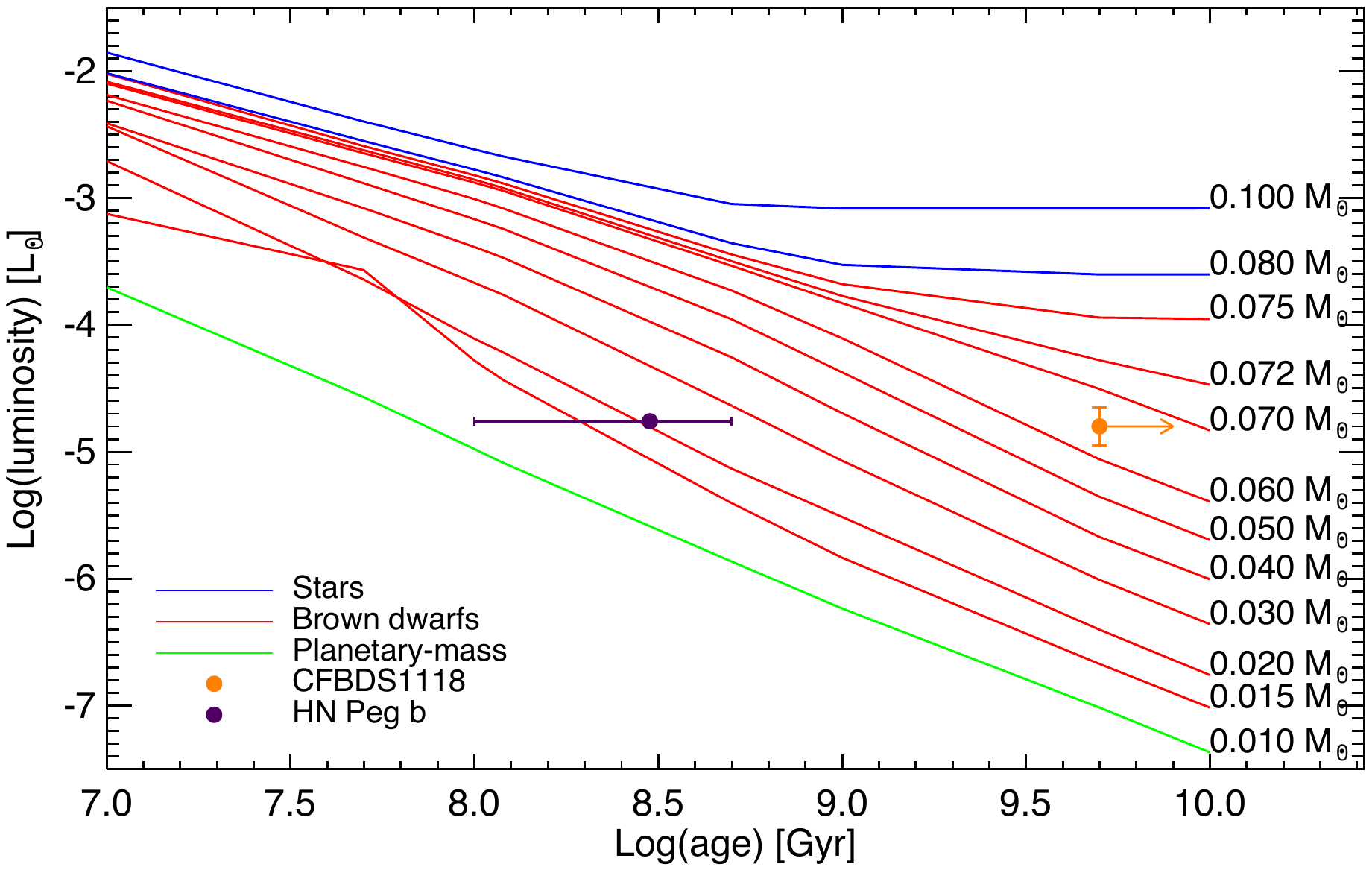}
\caption{
Luminosity of solar metallicity M dwarfs and substellar objects
against time after formation, from \cite{Baraffe.2003}. The black dot
marks \object{HN~Peg~B} and the orange dot shows \object{CFBDS~1118}. 
The luminosity of \object{HN~Peg~B} is log L/L$_\odot = -4.76$, with an error bar of 0.02 \citep{Leggett.2008},  smaller than the symbol size.
}
\label{burrows}
\end{center}
\end{figure}

Fig.~\ref{burrows} shows luminosity against time for different masses
for \cite{Baraffe.2003} models. The orange dot marks the position
of \object{CFBDS~1118}. The masses of old ($>$~5 Gyr) late-L and
  early-T dwarfs are much better constrained, for a given uncertainty 
  on their luminosity, than those of most sub-stellar objects. At such
  ages, objects below $\sim$ 60 Jupiter masses have cooled into
  late-Ts and cooler, while objects above $\sim$~78 Jupiter masses 
  remain on the bottom of the main sequence and are at most mid-Ls. 
  As a result, old age and an L-T transition spectral type, alone,
  constrain the mass to $\pm$15\%. Adding in an even a moderately 
  precise luminosity and an age pinpoints the mass, within the
  possible systematic errors in the theoretical model grid.  
  As an example, a 20\% uncertainty on a L $=10^{-4.5}$ L$_{\odot}$ luminosity 
  translates into a 3\% fractional mass uncertainty at 10 Gyr
  (74.9 $\pm$ 2.3 $M_{\rm{Jup}}$), and into a 13\% one at 0.5 Gyr 
  (36 $\pm$ 5.6  $M_{\rm{Jup}}$). Given our confidence in it old age,
  the discrepancies in the physical parameters determined from evolution
  and atmosphere models of \object{CFBDS~1118} are well established and make 
  an interesting testbed.

\section{Summary}
Brown dwarfs companions to main sequence stars provide unique
benchmarks for studies of cool atmospheres, because the
better understood primary star provides the age, distance, and 
metallicity, of the substellar companion. These systems therefore 
provide unique tests of cool atmosphere models by vastly reducing 
the number of degrees of freedom.

We identified a 7.7$^{\prime\prime}-$separation common proper motion M dwarf/T dwarf 
binary within our CFBD Survey \citep{Delorme.2008b}. The red colours of
the companion identify it as an early-T dwarf, as confirmed 
by its spectroscopic follow-up. The high tangential velocity of the 
system and the low upper limit on the $H_\alpha$ feature in the spectrum
of the M4.5-M5 primary indicates that the system is old 
($\gtrsim$ 5 Gyr). Observations of both components with 
the wide spectral coverage of XSHOOTER constrain the physical 
parameters of the system.

The T dwarf, with its metallicity fixed to [Fe/H] =
  $-0.1\pm0.1$ from the M dwarf and a well constrained mass 
estimate, provides a valuable anchor. Thanks to its older age, 
its properties depend only weakly on its exact age: since 
older brown dwarfs cool down and contract very slowly (the $T_{\rm{eff}}$ 
of a 70 $M_{\rm{Jup}}$ object, for instance, only drops from 1520 K 
to 1290 K between 5 and 10 Gyr, and its $\mathrm{log}\,g$ increase
almost insignificantly from 5.47 to 5.50), they are much
less affected by the age/mass/luminosity degeneracy that 
generally hinders field brown studies. The T dwarf also lies 
close to the boundary between the lowest mass stars (whose luminosity 
remains constant) and brown dwarfs (with decreasing luminosity). 
It is one of the most massive brown dwarfs, just below the hydrogen 
burning limit. 

The early-T spectral type of the brown dwarf enhances the value of 
the system, because the cloud-clearing that occurs at the L/T 
transition is very sensitive to gravity and metallicity 
\citep[e.g.][]{Oppenheimer.2013,Allers.2013}. This produces large scatter in the colours, with low gravity objects -- such as the late-L type planet 2M1207b \citep{Chauvin.2004} -- usually having redder J-K colours. Comparison with \object{HN~Peg~B}, 
a much younger T2.5 dwarf companion to a main sequence star, however,
finds that they have very similar photometric and
  spectroscopic properties. By contrast, the photometric properties 
of the two early-T components of \object{SIMP~1619+0313AB} system show
that they had very different dust dissipation histories in spite
of being coevals objects.
This suggests that the dust content of the atmosphere, which
  appears responsible for most of the scatter in L/T transition
  objects, is not simply controlled by gravity and metallicity: 
  though dust content generally increases with lower gravity and
  higher metallicity, several exceptions shows that additional
  parameters are important.

\begin{acknowledgements}
We acknowledge observing support by the CFHT and ESO staffs, and
thank the CFHT executive director for granting us discretionary
observing time. We acknowledge financial support from "Programme 
National de Physique Stellaire" (PNPS) of CNRS/INSU, France.
\end{acknowledgements}

\bibliographystyle{aa}
\bibliography{ref}

\end{document}